\documentclass[runningheads]{llncs}
\usepackage{algorithm}
\usepackage[noend]{algorithmic}
\usepackage{amssymb}
\usepackage{graphicx}
\usepackage{subfigure}
\usepackage{url}
\urldef{\mailsa}\path|{chihya, mschen}@citi.sinica.edu.tw,|
\urldef{\mailsb}\path|dnyang@iis.sinica.edu.tw,|
\urldef{\mailsc}\path|wlee@cse.psu.edu | 

\begin{document}

\title{Maximizing Friend-Making Likelihood for Social Activity Organization}
\author{Chih-Ya Shen$^1$%
\and De-Nian Yang$^2$\and Wang-Chien Lee$^3$\and Ming-Syan Chen$^{1,2}$}
\institute{Research Center for Information Technology Innovation, Academia Sinica \and Institute of Information Science, Academia Sinica\and Department of Computer Science and Engineering, The Pennsylvania State University\\
\mailsa\\
\mailsb
\mailsc\\}
\authorrunning{C.-Y. Shen, D.-N. Yang, W.-C. Lee, and M.-S. Chen}
\maketitle

\begin{abstract}
\vspace{-25pt} The social presence theory in social psychology suggests that
computer-mediated online interactions are inferior to face-to-face,
in-person interactions. In this paper, we consider the scenarios of
organizing in person friend-making social activities via online social networks (OSNs)
and formulate a new research problem, namely, Hop-bounded Maximum Group
Friending (HMGF), by modeling both existing friendships and the likelihood
of new friend making. To find a set of attendees for socialization
activities, HMGF is unique and challenging due to the interplay of the group
size, the constraint on existing friendships and the objective function on
the likelihood of friend making. We prove that HMGF is NP-Hard, and no
approximation algorithm exists unless $P=NP$. We then
propose an error-bounded approximation algorithm to efficiently obtain the
solutions very close to the optimal solutions. We conduct a user study to
validate our problem formulation and perform extensive experiments on real
datasets to demonstrate the efficiency and effectiveness of our proposed
algorithm. \vspace{-6pt}
\end{abstract}

\section{Introduction}
\vspace{-8pt} With the popularity and accessibility of online social
networks (OSNs), e.g., Facebook, Meetup, and Skout\footnote{%
http://www.skout.com/}, more and more people initiate friend gatherings or
group activities via these OSNs. For example, more than 16 millions of
events are created on Facebook each month to organize various kinds of
activities\footnote{%
http://newsroom.fb.com/products/}, and more than 500 thousands of
face-to-face activities are initiated in Meetup\footnote{%
http://www.meetup.com/about/}. The activities organized via OSNs cover a
wide variety of purposes, e.g., friend gatherings, cocktail parties,
concerts, and marathon events. The wide spectrum of these activities shows
that OSNs have been widely used as a convenient means for initiating
real-life activities among friends.

On the other hand, to help users expand their circles of friends in the
cyberspace, friend recommendation services have been provided in OSNs to
suggest candidates to users who may likely become mutual friends in the
future. Many friend recommendation services employ link prediction
algorithms, e.g., \cite{KA06,LK07}, to analyze the features, similarity or
interaction patterns of users in order to derive potential future friendship
between some users. By leveraging the abundant information in OSNs, link
prediction algorithms show high accuracy for recommending online friends in
OSNs.

As social presence theory \cite{SWC76} in social psychology suggests,
computer-mediated online interactions are inferior to face-to-face,
in-person interactions, off-line friend-making activities may be
favorable to their on-line counterparts in cyberspace. Therefore, in this paper, we consider
the scenarios of organizing face-to-face friend-making activities via OSN
services. Notice that finding socially cohesive groups of participants is
essential for maintaining good atmosphere for the activity. Moreover, the
function of making new friends is also an important factor for the success
of social activities, e.g., assigning excursion groups in conferences,
inviting attendees to housewarming parties, etc. Thus, for organizing friend-making social
activities, both activity organization and friend recommendation services
are fundamental. However, there is a gap between existing activity
organization and friend recommendation services in OSNs for the scenarios
under consideration. Existing activity organization approaches focus on
extracting socially cohesive groups from OSNs based on certain cohesive
measures, density, diameter, of social networks or other constraints, e.g.,
time, spatial distance, and interests, of participants \cite%
{YCL11,YSL12,ZHX14,SYY14}. On the other hand, friend recommendation
services consider only the \textit{existing friendships} to recommend
potential new friends for an individual (rather than finding a group of
people for engaging friend-making). We argue that in addition to themes of
common interests, it is desirable to organize friend-making activities by
mixing the "potential friends", who may be interested in knowing each other
(as indicated by a link prediction algorithm), with existing friends (as
lubricators). To the best knowledge of the authors, the following two
important factors, 1) the existing friendship among attendees, and 2) the
potential friendship among attendees, have not been considered
simultaneously in existing activity organization services. To bridge the
gap, it is desirable to propose a new activity organization service that
carefully addresses these two factors at the same time.

In this paper, we aim to investigate the problem of selecting a set of
candidate attendees from the OSN by considering both the existing and
potential friendships among the attendees. To capture the two factors for
activity organization, we propose to include the likelihood of making new
friends in the social network. As such, we formulate a new research problem
to find groups with tight social relationships among existing friends and
potential friends (i.e., who are not friends yet). Specifically, we model
the social network in the OSN as a heterogeneous social graph $G=(V,E,R)$
with edge weight $w:R\rightarrow (0,1]$, where $V$ is the set of
individuals, $E$ is the set of \textit{friend edges}, and $R$ is the set of 
\textit{potential friend edges} (or potential edges for short). Here a
friend edge $(u,v)$ denotes that individuals $u$ and $v$ are mutual friends,
while a potential edge $[u^{\prime},v^{\prime}]$ indicates that individuals $%
u^{\prime}$ and $v^{\prime}$ are likely to become friends (the edge weight $%
w[u^{\prime},v^{\prime}]$ quantifies the likelihood). The potential edges
and the corresponding edge weights can be obtained by employing a link
prediction algorithm in friend recommendation.

Given a heterogeneous social graph $G=(V,E,R)$ as described above, we
formulate a new problem, namely, \textit{Hop-bounded Maximum Group Friending
(HMGF)}, to find a group that 1) maximizes the likelihood of making new
friends among the group, i.e., the group has the highest ratio of total
potential edge weight to group size, 2) ensures that the social tightness,
i.e., hop count on friend edges in $G$ between each pair of individuals is
small, and 3) is a sufficiently large group, i.e., too small a group may not
work well for socialization activities.

\begin{figure}[tp]
\centering
\subfigure[][Input Graph $G$.] {\  \includegraphics[scale=0.35] {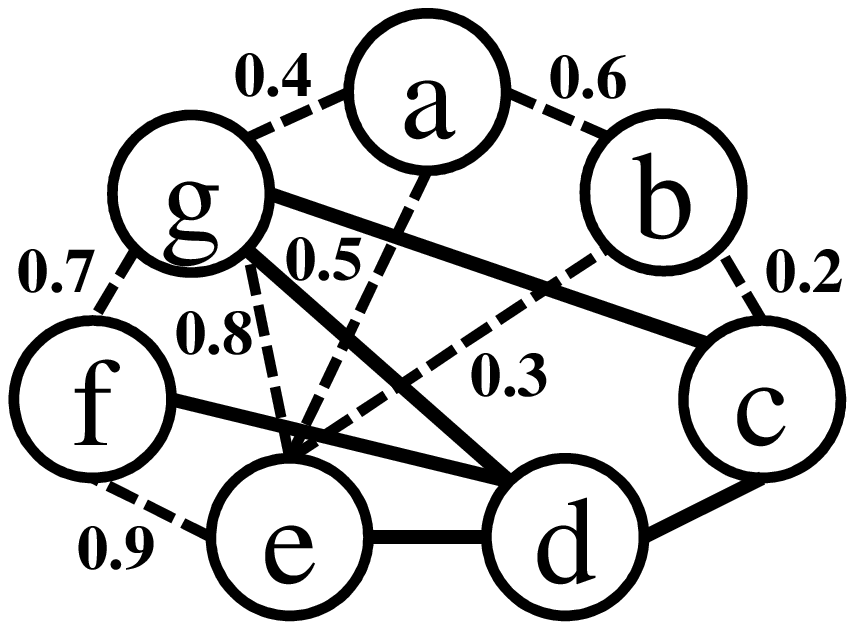}
\label{illu_exp} } 
\subfigure[][$H_1$.] {\  \includegraphics[scale=0.35] {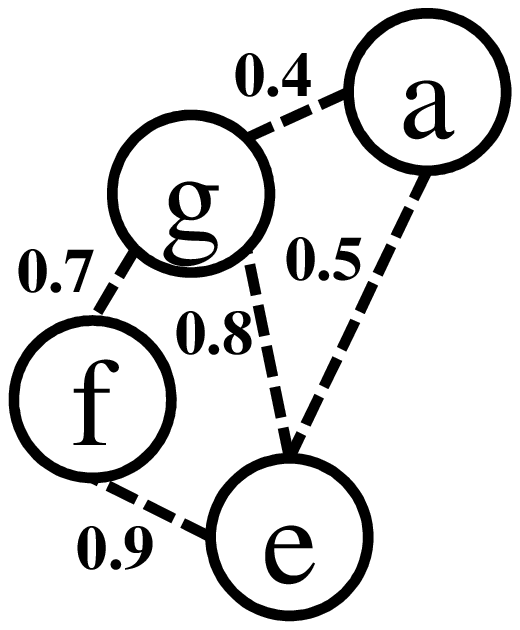}
\label{h_1} } 
\subfigure[][$H_2$.] {\  \includegraphics[scale=0.35] {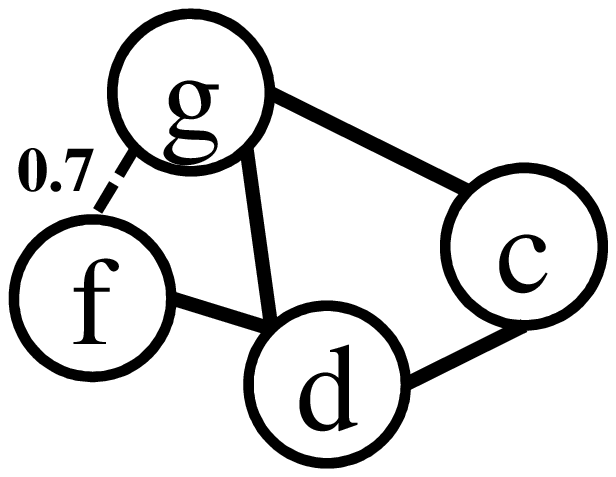}
\label{h_2} } 
\subfigure[][$H_3$.] {\  \includegraphics[scale=0.35] {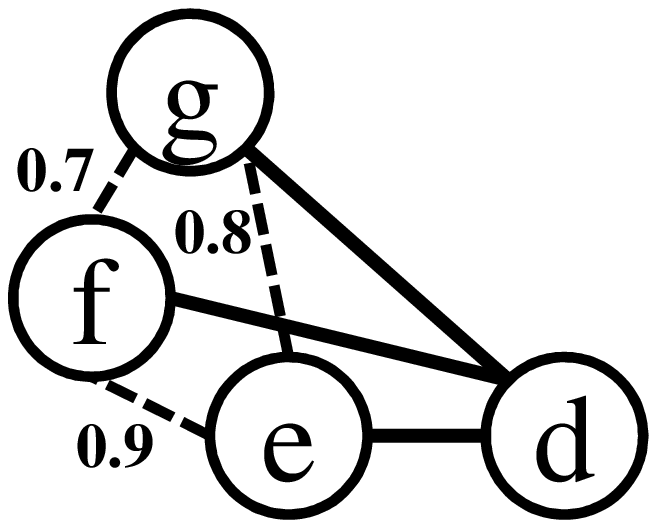}
\label{h_3} }
\par
\vspace{-10pt}
\caption{Illustrative Example.}
\vspace{-15pt}
\label{examples}
\end{figure}

Figure \ref{examples} illustrates the social graph and the interplay of the
above factors. Figure \ref{illu_exp} shows a social graph, where a dash
line, e.g., $[a,b]$ with weight 0.6, is a potential edge and a solid line,
e.g., $(c,d)$, is a friend edge. Figure \ref{h_1} shows a group $H_1$:$%
\{a,e,f,g\}$ which has many potential edges and thus a high total weight.
However, not all the members of this group have common friends as social
lubricators. Figure \ref{h_2} shows a group $H_2$:$\{c,d,f,g\}$ tightly
connected by friend edges. While $H_2$ may be a good choice for gathering of
close friends, the goal of friend-making in socialization activities is
missed. Finally, Figure \ref{h_3} shows $H_3$:$\{d,e,f,g\}$ which is a
better choice than $H_1$ and $H_2$ for socialization activities because each
member of $H_3$ is within 2 hops of another member via friend edges in $G$.
Moreover, the average potential edge weight among them is high, indicating
members are likely to make some new friends.

Processing HMGF to find the best solution is very challenging because there
are many important factors to consider, including hop constraint, group size
and the total weight of potential edges in a group. Indeed, we prove that
HMGF is an NP-Hard problem with no approximation algorithm.
Nevertheless, we prove that if the hop constraint can be slightly relaxed to allow a small
error, there exists a 3-approximation algorithm for HMGF.
Theoretical analysis and empirical results show that our algorithm can
obtain good solutions efficiently.

The contributions made in this study are summarized as follows.
\begin{itemize}
\item For socialization activity organization, we propose to model the
existing friendship and the potential friendship in a heterogeneous social
graph and formulate a new problem, namely, Hop-bounded Maximum Group
Friending (HMGF), for finding suitable attendees. To our best knowledge,
HMGF is the first problem that considers these two important relationships
between attendees for activity organization.

\item We prove that HMGF is NP-Hard and there exists no 
approximation algorithm for HMGF unless $P=NP$. We then propose an
approximation algorithm, called MaxGF, with a guaranteed error bound for
solving HMGF efficiently.

\item We conduct a user study on $50$ users to validate our argument for
considering both existing and potential friendships in activity
organization. We also perform extensive experiments on real datasets to
evaluate the proposed algorithm. Experimental results manifest that HMGF can
obtain solutions very close to the optimal ones, very efficiently.
\end{itemize}


\section{Problem Formulation}

\vspace{-8pt} \label{prob}

Based on the description of heterogeneous social graph described earlier,
here we formulate the \textit{Hop-bounded Maximum Group Friending (HMGF)}
tackled in this paper. Given two individuals $u$ and $v$, let $d_{G}^{E}(u,v)
$ be the shortest path between $u$ and $v$ via friend edges in $G$.
Moreover, given $H\subseteq G$, let $w(H)$ denote the total weight of
potential edges in $H$ and let \textit{average weight}, $\sigma (H)=\frac{w(H)}{%
|H|}$ denote the average weight of potential edges connected to each
individual in $H$\footnote{%
Note that $\sigma (H)=0$ if $H=\varnothing$.}. HMGF is formulated as follows.

\vspace{3pt} \noindent \textbf{Problem: Hop-bounded Maximum Group Friending
(HMGF).}

\noindent \textbf{Given: } Social network $G=(V,E,R)$, hop constraint $h$,
and size constraint $p$.

\noindent \textbf{Objective: } Find an induced subgraph $H\subseteq G$ with
the maximum $\sigma (H)$, where $|H|\geq p$ and $d_{G}^{E}(u,v)\leq h,\forall
u,v\in H$.

Efficient processing of HMGF is very challenging due to the following
reasons: 1) The interplay of the total weight $w(H)$ and the size of $H$. To
maximize $\sigma (H)$, finding a small $H$ may not be a good choice because
the number of edges in a small graph tends to be small as well. On the other
hand, finding a large $H$ (which usually has a high $w(H)$) may not lead to
an acceptable $\sigma (H)$, either. Therefore, the key is to strike a good
balance between the graph size $|H|$ and the total weight $w(H)$. 2) HMGF
includes a hop constraint (say $h=2$) on friend edges to ensure that every
pair of individuals is not too distant socially from each other. However, 
selecting a potential edge $[u,v]$ with a large weight $w[u,v]$ may not necessarily satisfy
the hop constraint, i.e., $d_{G}^{E}(u,v)> h$ which is defined based on existing friend edges. 
In this case, it may not always be a good strategy to prioritize on large-weight edges in order to maximize $\sigma (H)$,
especially when $u$ and $v$ do not share a common friend nearby via the friend
edges.

In the following, we prove that HMGF is NP-Hard and \textit{not approximable}
within any factor. In other words, there exists no approximation algorithm
for HMGF.

\begin{theorem}
\label{thm_np} HMGF is NP-Hard and there is no approximation algorithm for HMGF unless $P=NP$.
\end{theorem}

\begin{proof}
Due to the space constraints, we prove this theorem in the full version of this paper (available online \cite{fullversion}). 
\end{proof}


\vspace{-8pt}

\section{Related Work}

\label{relatedwork} \vspace{-8pt} Extracting dense subgraphs or social
cohesive groups among social networks is a natural way for selecting a set
of close friends for a gathering. Various social cohesive measures have been
proposed for finding dense social subgraphs, e.g., diameter \cite{WF94}, density \cite{FS97},
clique and its variations \cite{M79}. Although these
social cohesive measures cover a wide range of application scenarios, they
focus on deriving groups based only on existing friendship in the social
network. In contrast, the HMGF studied in this paper aims to extract groups
by considering both the existing and potential friendships for socialization
activities. Therefore, the existing works mentioned above cannot be directly
applied to HMGF tackled in this paper.

Research on finding a set of attendees for activities based on the
social tightness among existing friends \cite%
{YCL11,YSL12,ZHX14,SYY14,SLL11} have been reported in the literature.
Social-Temporal Group Query 
\cite{YCL11} checks the available times of attendees to find the social
cohesive group with the most suitable activity time. Geo-Social Group Query 
\cite{YSL12,ZHX14} extracts socially tight groups while considering
certain spatial properties. The willingness optimization for social
group problem in \cite{SYY14} selects a set of attendees for an activity
while maximizing their willingness to participate. Finally, \cite{SLL11}
finds a set of compatible members with tight social relationships in the collaboration network.
Although these works find
suitable attendees for activities based on existing friendship
among the attendees, they ignore the likelihood of making new friends among
the attendees. Therefore, these works may not be suitable for socialization
activities discussed in this paper.

Link prediction analyzes the features,
similarity or interaction patterns among individuals in order to recommend
possible friends to the users \cite{KA06,LK07,CMN08,KL09,LLL10}. Link prediction algorithms employ different approaches
including graph-topological features, classification models, hierarchical probabilistic model, and
linear algebraic methods. These works show good prediction
accuracy for friend recommendation in social networks. In this paper, to
estimate the likelihood of how individuals may potentially become friends in
the future, we employ link prediction algorithms for deriving the potential
edges among the individuals.

To the best knowledge of the authors, there exists no algorithm for activity organization
that considers both the existing friendship and the likelihood of
making new friends when selecting activity attendees. The HMGF studied in
this paper examines the social tightness among existing friends and the
likelihood of becoming friends for non-friend attendees. We envisage that
our research result can be employed in various social network applications
for activity organization.

\section{Experimental Results}

\label{exp} \vspace{-8pt} We implement HMGF in Facebook and invite 50 users
to participate in our user study. Each user, given 12 test cases of HMGF
using her friends in Facebook as the input graph, is asked to solve the HMGF
cases, and compare her results with the solutions obtained by MaxGF. In
addition to the user study, we evaluate the performance of MaxGF on two
real social network datasets, i.e., FB \cite{VMM09} and the MS dataset from
KDD Cup 2013\footnote{%
https://www.kaggle.com/c/kdd-cup-2013-author-paper-identification-challenge/data%
}. The FB dataset is extracted from Facebook with 90K vertices, and MS is a
co-author network with 1.7M vertices. We extract the friend edges from these
datasets and identify the potential edges with a link prediction algorithm 
\cite{LK07}. The weight of a potential edge is ranged within (0,1].
Moreover, we compare MaxGF with two algorithms, namely, Baseline and DkS 
\cite{FS97}. Baseline finds the optimal solution of HMGF by enumerating all
the subgraphs satisfying the constraints, while DkS is an $O(|V|^{1/3})$%
-approximation algorithm for finding a $p$-vertex subgraph $H\subseteq G$
with the maximum density on $E\cup R$ without considering the potential
edges and the hop constraint. The algorithms are implemented in an IBM 3650
server with Quadcore Intel X5450 3.0 GHz CPUs. We measure 30 samples in each
scenario. In the following, FeaRatio and ObjRatio respectively denote the
ratio of feasibility (i.e., the portion of solutions satisfying the hop
constraint) and the ratio of $\sigma(H)$ in the solutions obtained by MaxGF
or DkS to that of the optimal solution.

\vspace{-4pt}

\subsection{User Study}

\vspace{-4pt} 
\begin{figure}[t]
\centering
\subfigure[][Required Time.] {\  \includegraphics[scale=0.15] {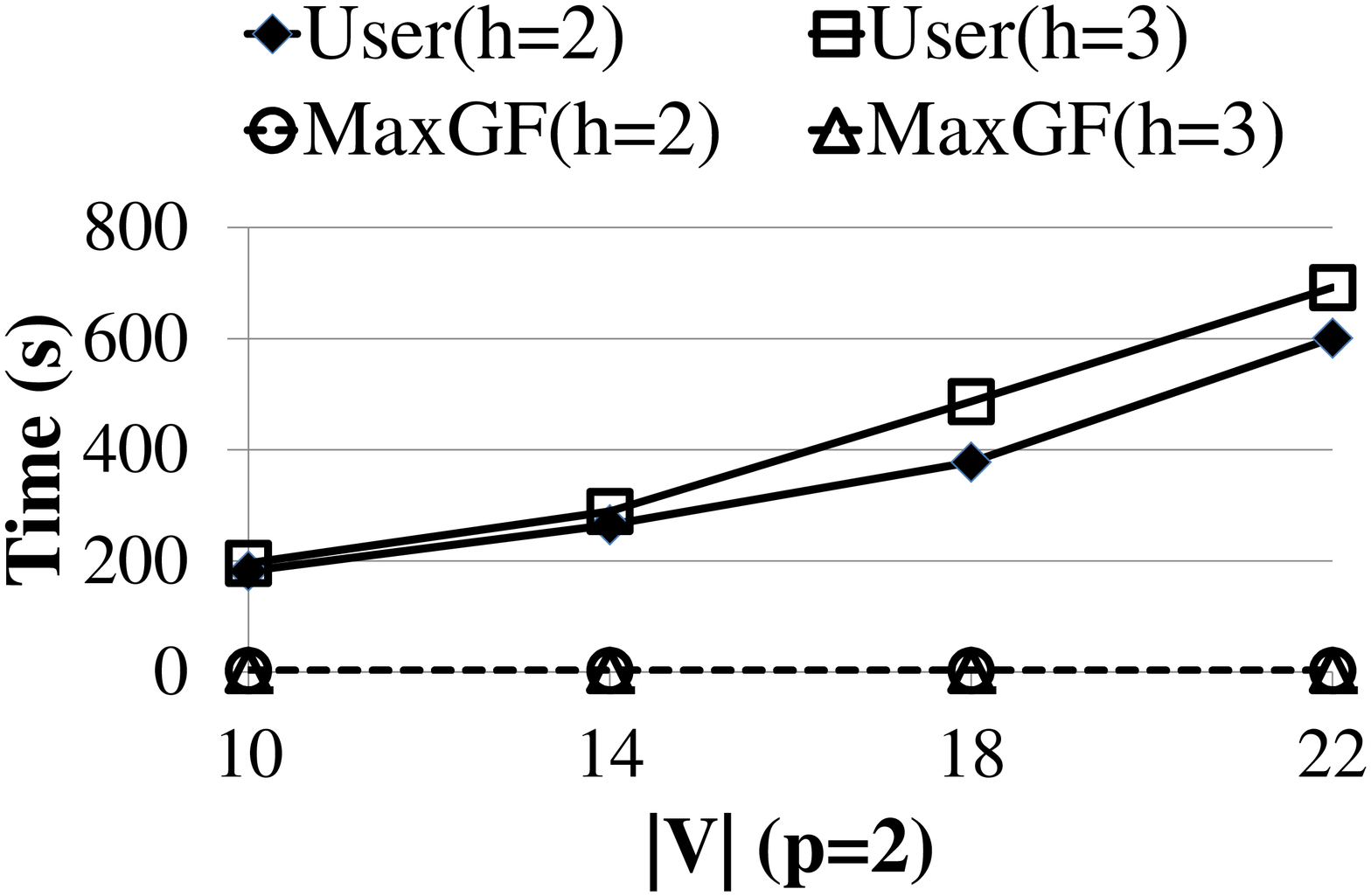}
\label{us_time} } 
\subfigure[][FeaRatio and ObjRatio.] {\  \includegraphics[scale=0.15] {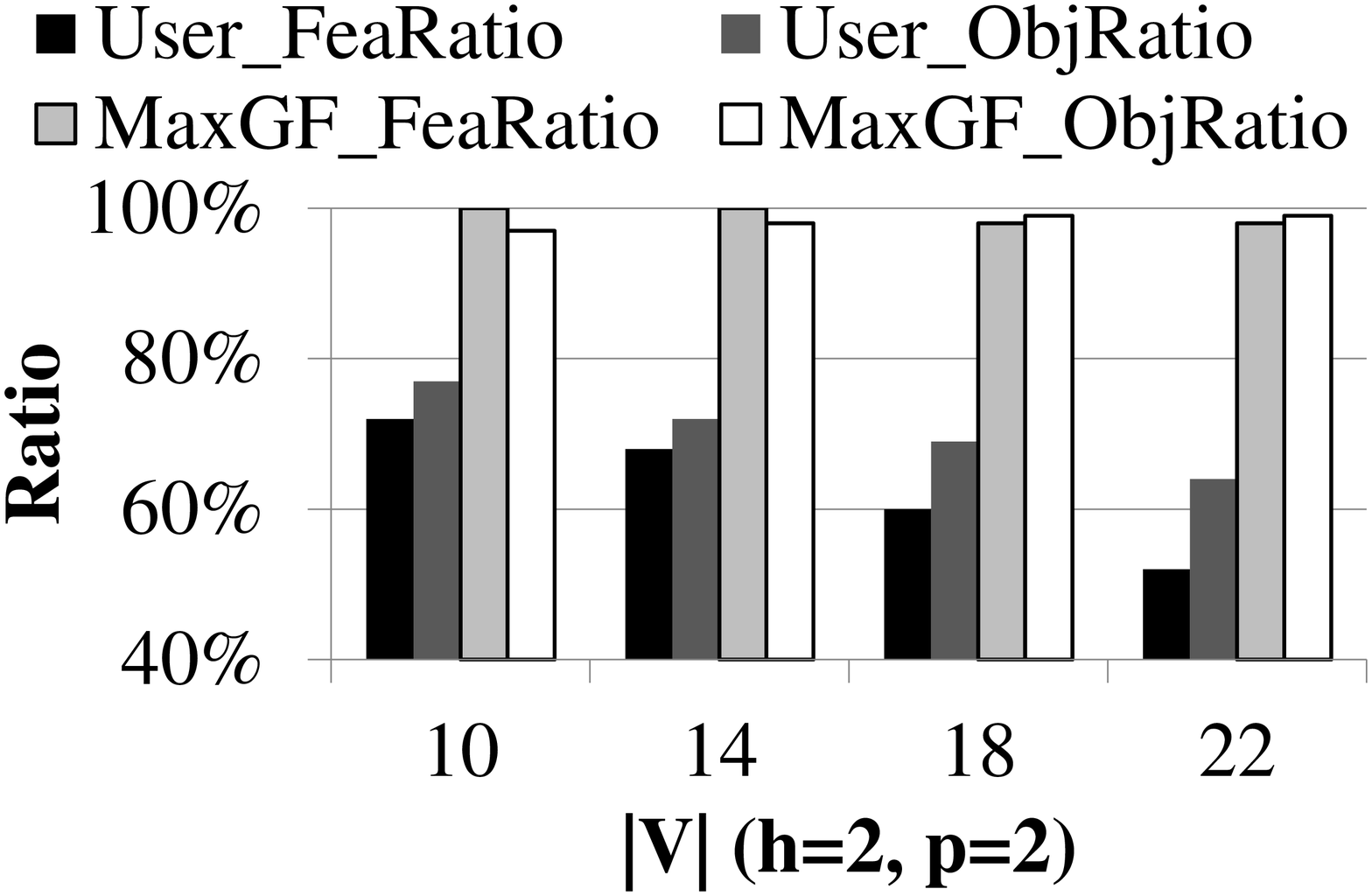}
\label{us_ratio} } 
\subfigure[][User Satisfaction.] {\  \includegraphics[scale=0.15] {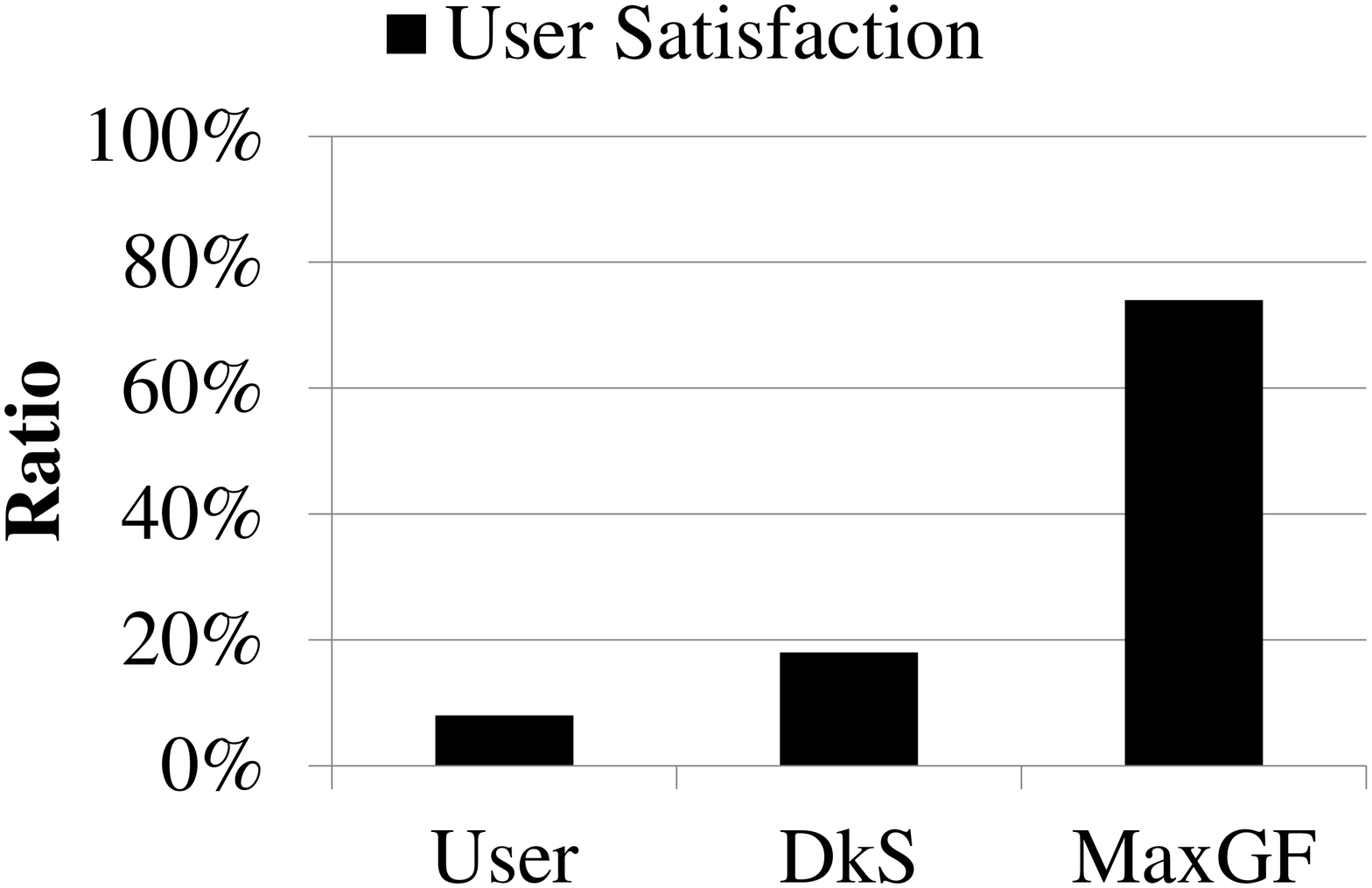}
\label{us_satisfaction} } 
\vspace{-10pt}
\caption{User Study Results.}
\vspace{-5pt}
\label{exp_us}
\end{figure}

Figure \ref{exp_us} presents the results of the user study. Figure \ref%
{us_time} compares the required time for users and MaxGF to solve the HMGF
instances. Users need much more time than MaxGF due to challenges brought
by the hop constraint and tradeoffs in potential edge weights and the group
size, as explained in Section \ref{prob}. As $|V|$ or $h$ grows, users need
more time because the HMGF cases become more complicated.
Figure \ref{us_ratio} compares the solution feasibility and quality among
users and MaxGF. We employ Baseline to obtain the optimal solutions and
derive FeaRatio and ObjRatio accordingly. The FeaRatio and ObjRatio of users
are low because simultaneously considering both the hop constraint on friend
edges and total weights on potential edges is difficult for users. As shown,
users' FeaRatio and ObjRatio drop when $|V|$ increases. By contrast, MaxGF
obtains the solutions with high FeaRatio and ObjRatio. In Figure \ref%
{us_satisfaction}, we ask each user to compare her solutions with the
solutions obtained by MaxGF and DkS, to validate the effectiveness of HMGF.
74\% of the users agree that the solution of MaxGF is the best because HMGF
maximizes the likelihood of friend-making while considering the hop
constraint on friend edges at the same time. By contrast, DkS finds the
solutions with a large number of edges, but it does not differentiate the
friend edges and potential edges. Therefore, users believe that the selected
individuals may not be able to socialize with each other effectively.

\vspace{-4pt}

\subsection{Performance Evaluation}

\vspace{-4pt} 
\begin{figure}[t]
\centering
\subfigure[][Time of Diff. $|V|$.] {\  \includegraphics[scale=0.15] {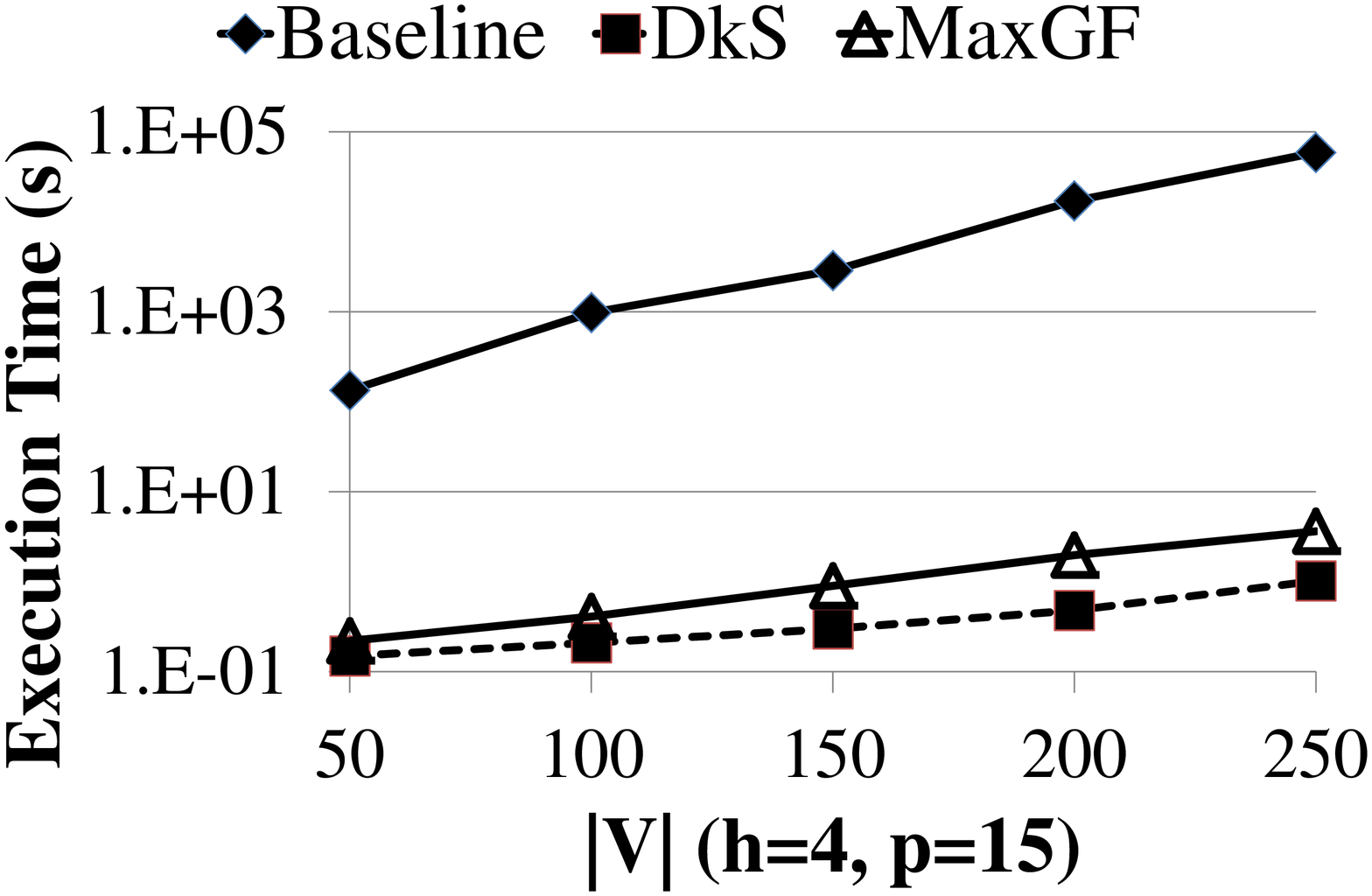}
\label{baseline_time_V} } 
\subfigure[][FeaRatio of Diff. $|V|$.] {\  \includegraphics[scale=0.15] {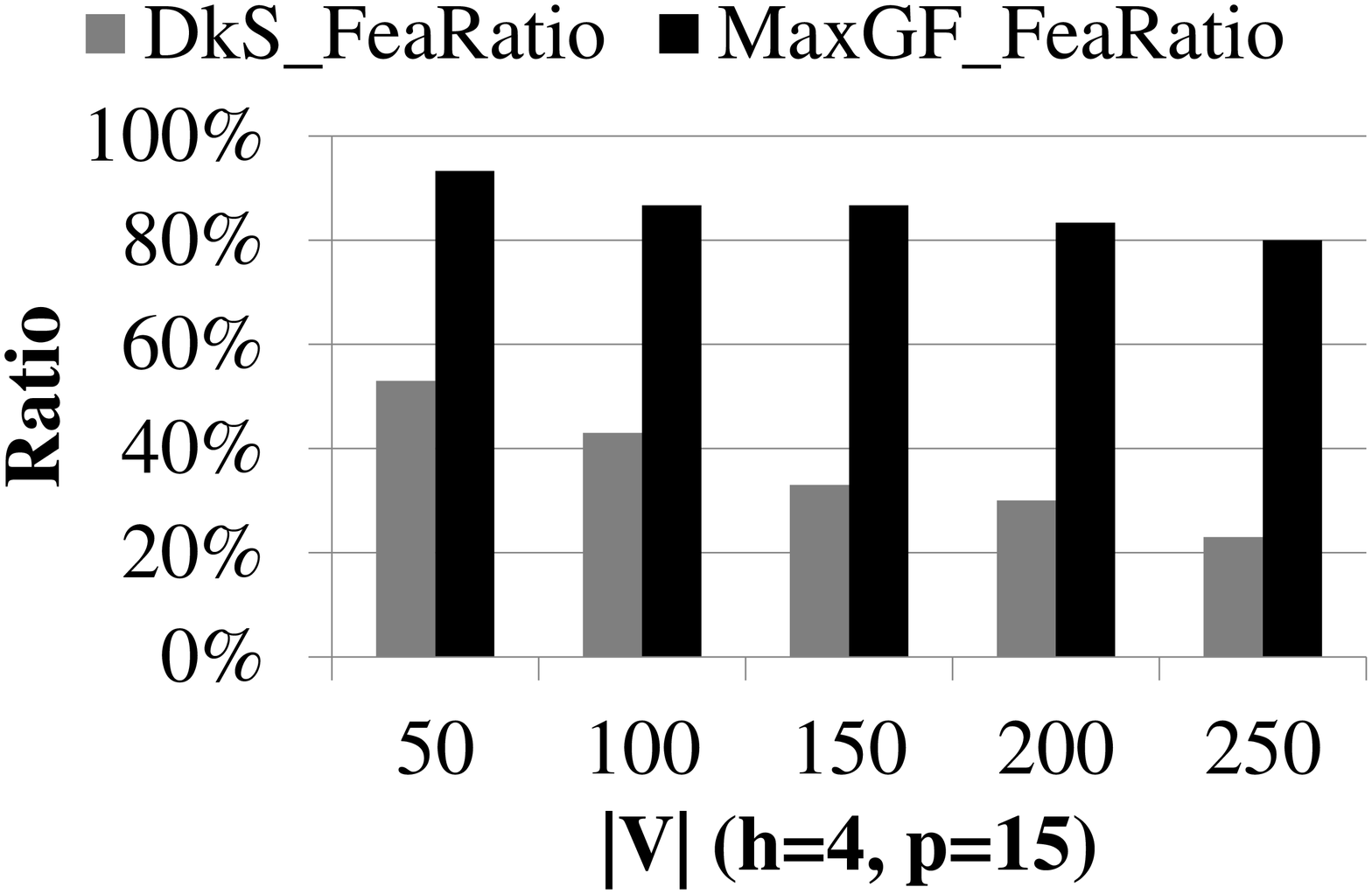}
\label{baseline_fea_V} } 
\subfigure[][ObjRatio of Diff $|V|$.] {\  \includegraphics[scale=0.15] {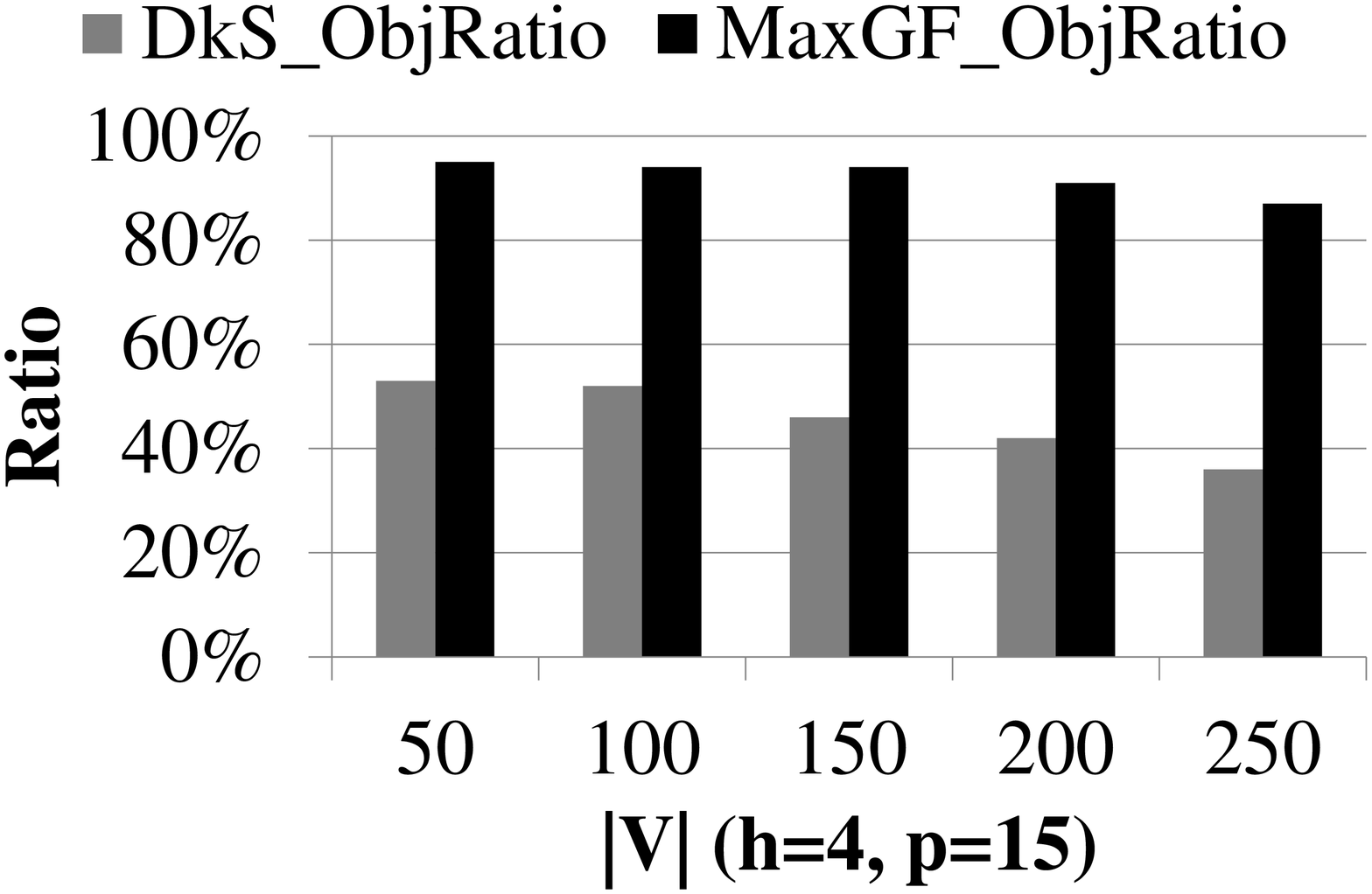}
\label{baseline_obj_V} } 
\subfigure[][Time of Diff. $h$.] {\  \includegraphics[scale=0.15] {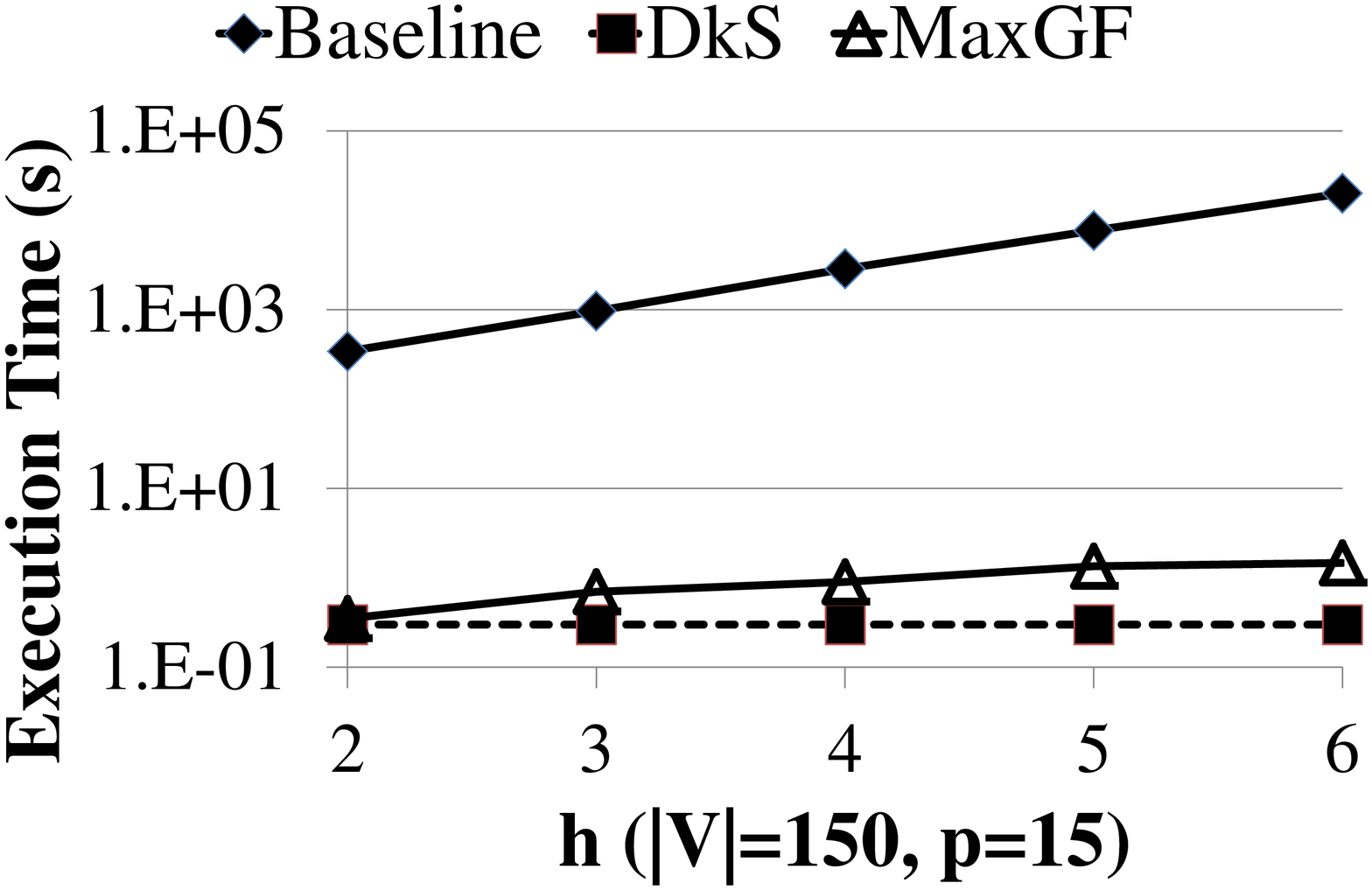}
\label{baseline_time_h} } 
\subfigure[][FeaRatio of Diff. $h$.] {\  \includegraphics[scale=0.15] {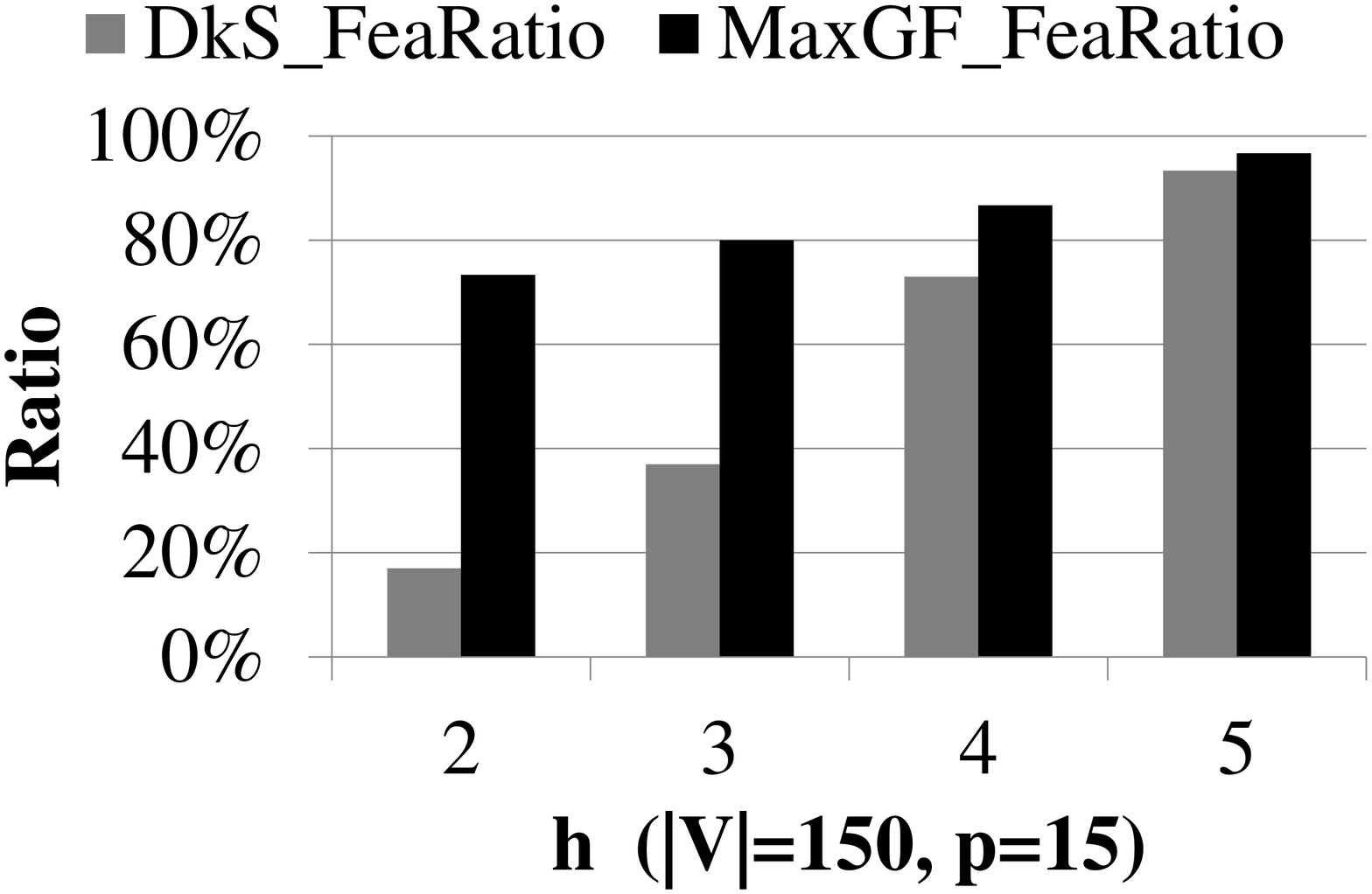}
\label{baseline_fea_h} } 
\subfigure[][ObjRatio of Diff. $h$.] {\  \includegraphics[scale=0.15] {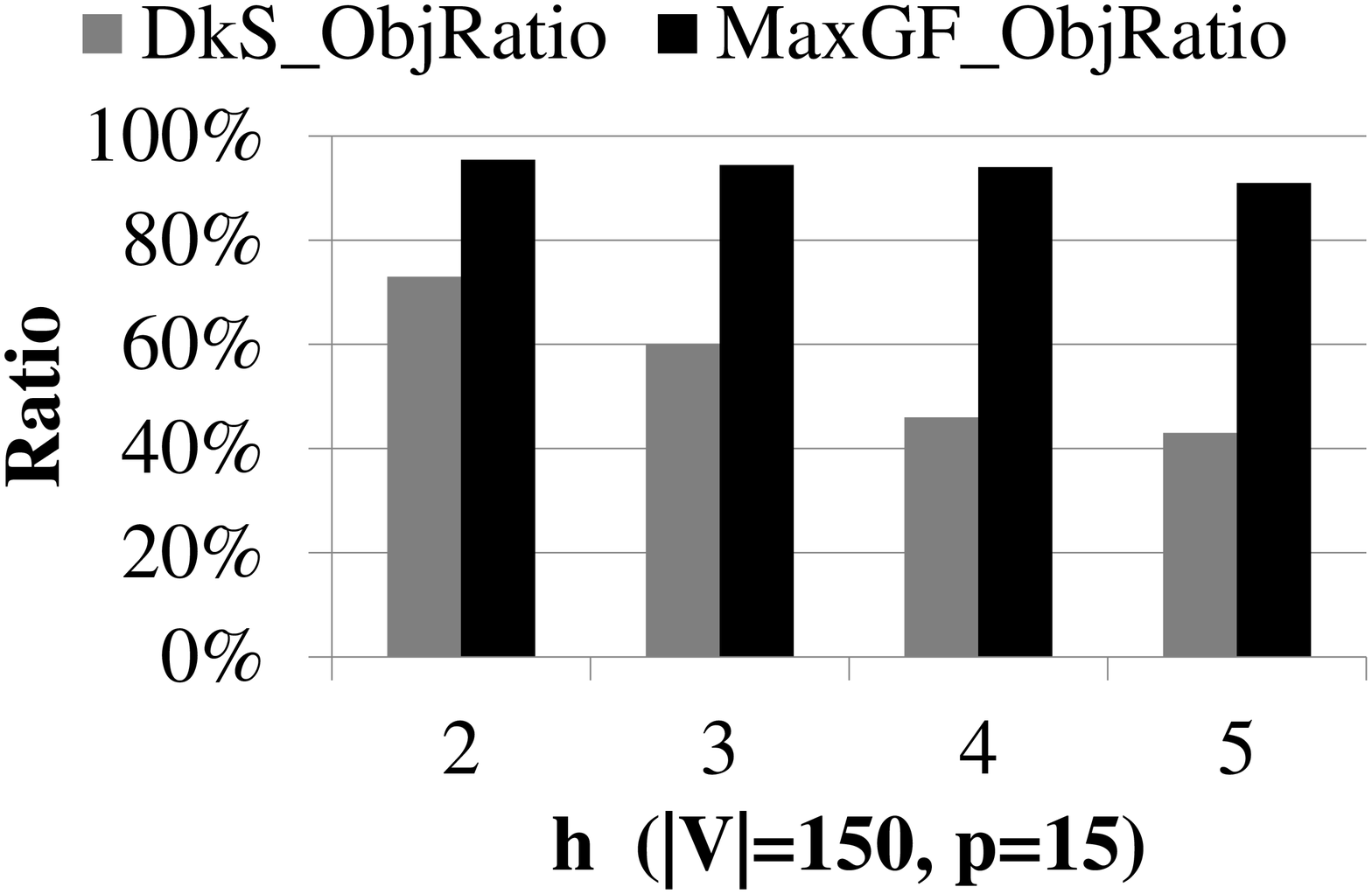}
\label{baseline_obj_h} } \vspace{-8pt}
\caption{Comparisons with Optimal Solutions.}
\vspace{-10pt}
\label{exp_baseline}
\end{figure}

Baseline can only find the optimal solutions of small HMGF cases since it
enumerates all possible solutions. Therefore, we first compare MaxGF
against Baseline and DkS on small graphs randomly extracted from FB. Figure %
\ref{baseline_time_V} compares the execution time of the algorithms by
varying the size of input graph. Since Baseline enumerates all the subgraphs 
$H$ with $|H|\geq p$, the execution time grows exponentially. The execution
time of MaxGF is very small because the hop-bounded subgraphs and the
pruning strategy effectively trim the search space. Figures \ref%
{baseline_fea_V} and \ref{baseline_obj_V} present the FeaRatio and ObjRatio
of the algorithms, respectively. MaxGF has high ObjRatio because MaxGF
iteratively removes vertices with low incident weights from each hop-bounded
subgraph $H_{v}$, and extracts the solution $S^{APX}$ with maximized $\sigma
(S^{APX})$ among different subgraphs in different $H_{v}$ to strike a good
balance on total edge weights and group sizes as describe in Section \ref{prob}. Moreover, the high FeaRatio
and ObjRatio also indicate that the post-processing procedure effectively
restores the hop constraint and maximizes the average weight accordingly. By
contrast, DkS does not consider the hop constraint and different edge types
in finding solutions and thus generates the solutions with smaller FeaRatio
and ObjRatio.

\begin{figure}[tp]
\centering
\subfigure[][FeaRatio of Diff. $h$.] {\  \includegraphics[scale=0.15] {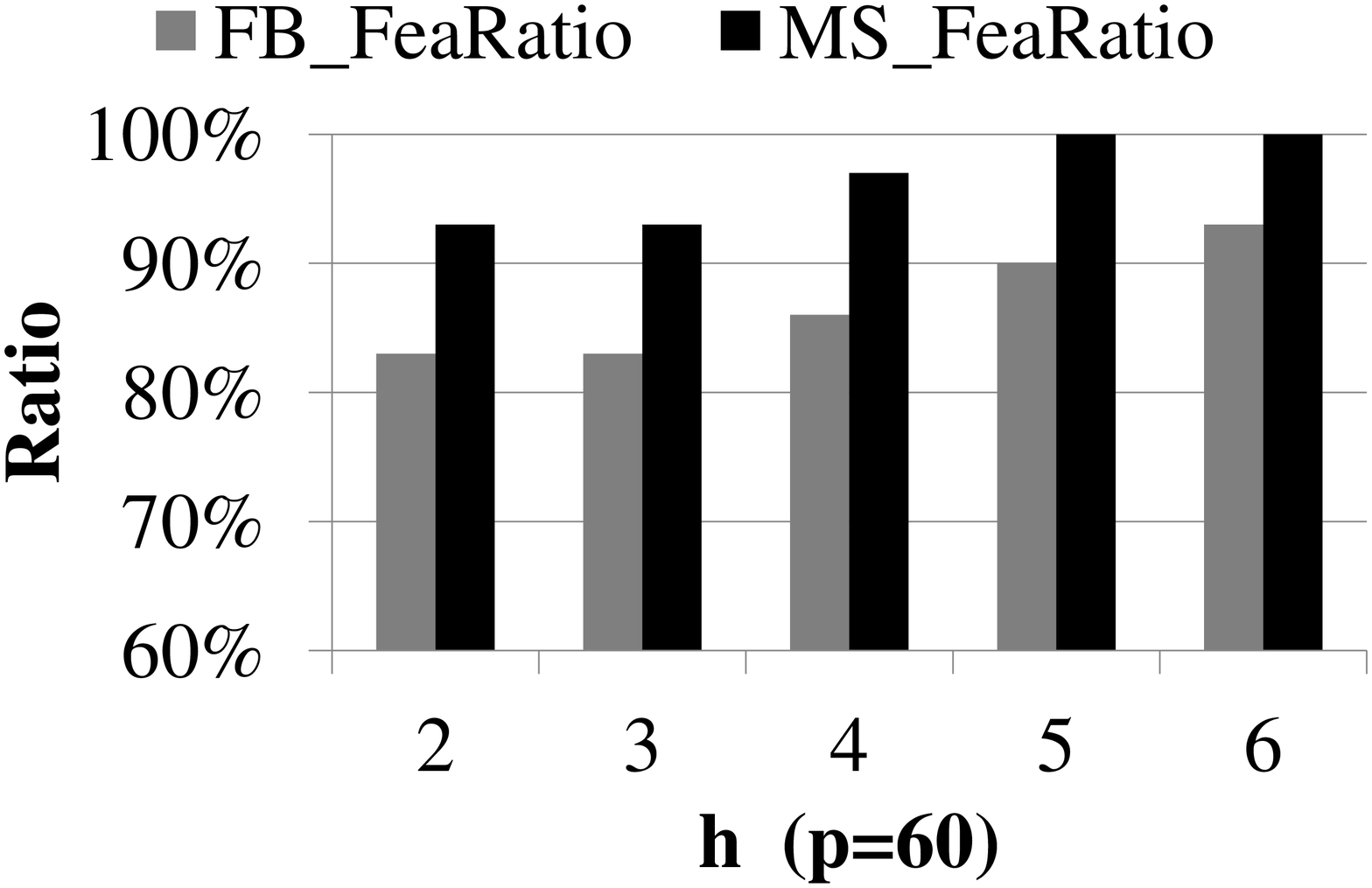}
\label{fea_H} } 
\subfigure[][$|S^{APX}|$ of Diff. $h$.] {\  \includegraphics[scale=0.15] {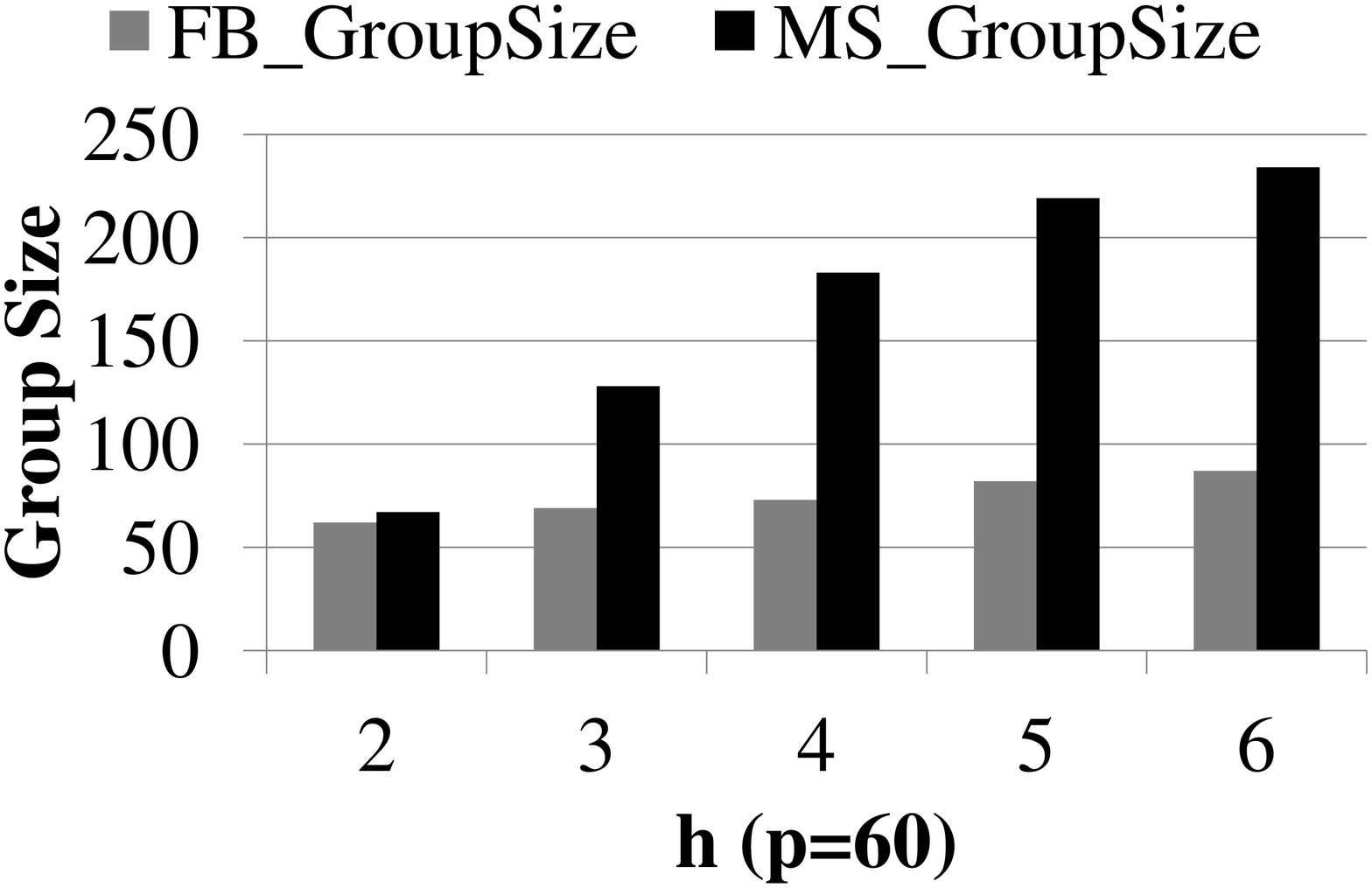}
\label{group_H} } 
\subfigure[][Time of Diff. $p$.] {\  \includegraphics[scale=0.15] {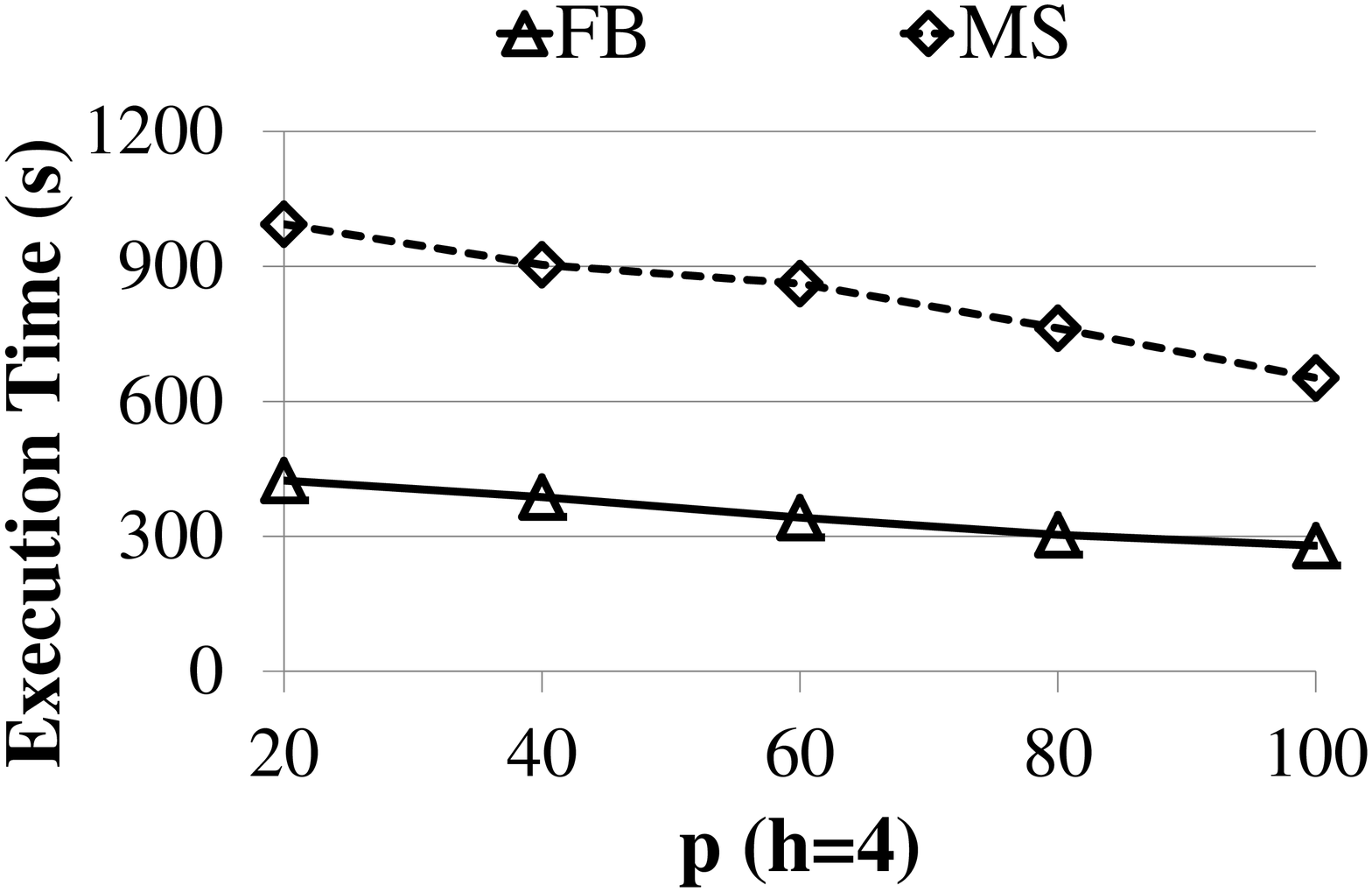}
\label{time_P} } \vspace{-10pt}
\caption{Experimental Results on Different Datasets.}
\vspace{-10pt}
\label{exp_datasets}
\end{figure}

Figures \ref{baseline_time_h}-(f) compare execution time, FeaRatio and
ObjRatio again but by varying $h$. When $h$ increases, the execution time of
MaxGF grows slowly because the pruning strategy avoids examining the
hop-bounded subgraphs that do not lead to a better solution. The FeaRatio
and ObjRatio of MaxGF with different $h$ are high because MaxGF employs
hop-bounded subgraphs to avoid generating solutions with large hop distances
on friend edges, and the post-processing procedure effectively restores the
hop constraint and maximizes the objective function.

Figure \ref{exp_datasets} compares MaxGF in different datasets, i.e., FB
and MS. 
Figures \ref{fea_H} and \ref{group_H} present the FeaRatio and the solution
group sizes with different $h$. As $h$ increases, MaxGF on both datasets
achieves a higher FeaRatio due to the post-processing procedure adjusts $%
S^{APX}$ and further minimizes $d_{G}^{E}(u,v),\forall u,v\in S^{APX}$.
Moreover, it is worth noting that the returned group sizes grow when $h$
increases in MS. This is because MS contains large densely connected
components with large edge weights. When $h$ is larger, MaxGF is inclined
to extract larger groups from these components to maximize the objective
function. By contrast, FB does not have large components and MaxGF thereby
tends to find small groups to reduce the group size for maximizing the
objective function. In fact, the solutions in FB are almost the same with
different $h$. Finally, MaxGF needs to carefully examine possible solutions
with the sizes at least $p$, and thus Figure \ref{time_P} shows that when $p$
increases, the execution time drops because MaxGF effectively avoids
examining the candidate solutions with small group sizes.

\vspace{-10pt}

\section{Conclusion}

\label{conclu} \vspace{-8pt} To bridge the gap between the state-of-the-art
activity organization and friend recommendation in OSNs, in this
paper, we propose to model the individuals with existing and potential
friendships in OSNs for friend-making activity organization. We
formulate a new research problem, namely, Hop-bonded Maximum Group Friending (HMGF), 
to find suitable activity attendees. We prove that
HMGF is NP-Hard and there exists no approximation algorithms unless $P=NP$.
We then propose an approximation algorithm with guaranteed error bound,
i.e., MaxGF, to find good solutions efficiently. We conduct a user study
and extensive experiments to evaluate the performance of MaxGF, where
MaxGF outperforms other relevant approaches in both solution quality and
efficiency.

\end{document}